\documentclass[aps,prl,twocolumn,superscriptaddress]{revtex4}
\usepackage{amsfonts}
\usepackage{amsmath}
\usepackage{amssymb}
\usepackage{graphicx}
\setcitestyle{super}

\begin{document}

\title{How does the motion of the surrounding molecules depend on the shape of a folding molecular motor ?}


\author{Simona Ciobotarescu}
\affiliation{ Laboratoire de Photonique d'Angers EA 4464, University of Angers, Physics Department,  2 Bd Lavoisier, 49045 Angers, France.}
\affiliation{Gheorghe Asachi Technical University of Iasi, Department of Natural and Synthetic Polymers, 73, Prof. Dimitrie. Mangeron Street, 700050 Iasi, Romania.}

\author{Nicolae Hurduc}
\affiliation{Gheorghe Asachi Technical University of Iasi, Department of Natural and Synthetic Polymers, 73, Prof. Dimitrie. Mangeron Street, 700050 Iasi, Romania.}

\author{Victor Teboul }
\email{victor.teboul@univ-angers.fr}
\affiliation{ Laboratoire de Photonique d'Angers EA 4464, University of Angers, Physics Department,  2 Bd Lavoisier, 49045 Angers, France.}


\begin{abstract}
Azobenzene based molecules have the property to isomerize when illuminated.  
In relation with that photoisomerization property, azobenzene containing materials are the subject of unexplained massive mass transport.
In this work we use an idealised rectangular chromophore model to study the dependence of the isomerization induced transport on the chromophore's dimensions. Our results show the presence of a motor arm length threshold for induced transport, which corresponds to the host molecule's size.
Above the threshold, the diffusive motions increase proportionally to the chromophore's length.
Intriguingly, we find only a very small chromophore width dependence of the induced diffusive motions.
Our very simplified motor reproduce relatively well the behavior observed using the real $DR_{1}$ motor molecule, suggesting that the complex closing procedure and detailed shape of the motor are not necessary to induce the molecular motions.

\end{abstract}

\maketitle


\footnotetext{$\ast$ E-mail: victor.teboul@univ-angers.fr}

\section{Introduction}

The design and understanding of molecular motors have attracted considerable attention\cite{motor1,motor2,motor3,motor4,motor5,motor6,motor7,motor8,motor10,motor11,motor12,motor13,motor14,motor15,motor16,motor17} since the development of nanotechnology that followed the visionary idea of R. Feynman.
While biological motor proteins are ubiquitous in living organisms, synthetic molecular motors are also of paramount interest due to their simpler mechanisms, robustness in various environments and smaller size.
Molecules like stilbene, azobenzene and their derivatives, due to their property of photo-isomerization are of particular interest as motors, because they do not consume or reject any waste inside the host medium and thus are able to continue their motion indefinitely provided that they are illuminated.

When illuminated, azobenzene doped materials are subject to intriguing massive mass transport.
While the physical mechanism leading to that transport is still a matter of debate\cite{review,review2,review4,a17,a18,a21,a22,a23,a24}, there is no doubt that its origin is to be found in the photo-isomerization property of the azobenzene molecule.
Recently several important results have improved our understanding of the mechanism.
i) It was reported\cite{diamond} that the pressure necessary to stop the isomerization of the molecule is very large ($P>1 GPa$).
ii) A fluidization or a softening of the host material around the chromophores was reported by different groups\cite{dif1,dif2,dif3,dif4}.
iii) If the massive mass transport is usually obtained from an interference pattern using two laser beams it was found\cite{regis,regis2,regis3} that even without this pattern, thus with only one beam, the effect took place.
iv) A large increase of the diffusive motions of host molecules around the chromophore during its isomerization has been reported in different works\cite{prl,cage} using molecular dynamics simulations. As the Stokes-Einstein relation  relates the diffusion to the viscosity, these results are in deep agreement with the experimental local fluidization that has been reported.

Molecular dynamics simulations\cite{md1,md2,md2b} (MD) permit us to gain information on the motion of each molecule of the medium without any  assumption on the origin of the unexplained isomerization-induced transport phenomenon. 
MD simulation is thus an invaluable tool\cite{md3,md4,md6,md7,md8,md9} to study the origin of the isomerization-induced molecular motions in azobenzene containing materials and more generally condensed matter physics phenomena.
Note however that MD simulations are limited to short time scales and cannot access the large time range needed to observe the appearance of the induced patterns.
Previous MD simulations found that the molecular motions increase near the motor while observing no other important effects. 
Note that it has been demonstrated analytically\cite{regis4} that an isomerization-induced increase in diffusive molecular motions is sufficient to explain the appearance of the induced patterns.
It is thus tempting to make the hypothesis that these induced molecular motions are at the origin of the massive mass transport resulting in the induced patterns that are observed experimentally for larger time scales.

In reality the picture is somewhat more complicated as we expect different physical mechanisms to appear sequentially during the surface relief gratings (SRG) formation\cite{review,a26}. For short time scales the chromophore's isomerization induces molecular rearrangements nearby that lead to the motion of surrounding host molecules\cite{cage,a22,a23,a24,prl} and to its own motion resulting in the rotation of the chromophore even at low temperatures when the thermal diffusion is small.  Then due to the preferential light absorption in the chromophore's dipole direction, the chromophores align themselves along a direction perpendicular to the electric field of the incident light\cite{review,a24} leading to the appearance of new physical mechanisms\cite{a17,a18,a19,a21}.
Recent experiments\cite{a26} show that the two types of physical mechanisms (dipoles induced or isomerization induced) cohabit also for larger time scales, a result that one expects as long as there are still isomerizations in the medium.
We are interested here only in the short timescale physical mechanisms that are directly induced by the isomerizations, as the effect of the alignment has already been studied extensively\cite{a17,a18,a19,a21,review}.

A number of experiments have been realized in the past decades with the aim of finding chromophores and host materials that maximize the massive mass transports that lead to the surface relief gratings (SRG) formation\cite{review}.
However 
 the main physical parameters that control the mass transport are still unknown. Molecular dynamics simulations have the advantage on experiments to permit simplifications of the system, leading to easier interpretations. As the temperature and density effects are now relatively well understood, the main remaining parameters that may control the molecular motions are the shape and relative size of the motor in comparison to the host.

In this work, using molecular dynamics simulations, we thus investigate the effect of the shape and relative size of the molecular motor on the induced transport. Our main purpose is to find the relevant physical parameters that control the transport mechanism and as a result the SRG formation, with the hope to lead to guidelines for experimentalists. We also expect from the study a better understanding of the physical phenomena at the origin of the massive mass transport.

To be able to extract the shape effect we simplify the motor molecule as much as possible and model the motor as flat with a rectangular shape in its relaxed (cis) form.
In the same spirit we model the motor's closure as a simple folding of the molecule around the center axis (see Figures 1 and 2).
We find that the simplifications modify  the motor's efficiency only slightly, a result that is in agreement with  the large number of azo dye molecules inducing surface relief gratings.  
This result shows that the exact particular shape and closure of the DR1 motor is not necessary to induce molecular motions  inside the host medium.
We study the shape dependence of the motor's efficiency and find the presence of thresholds on the motor's length while the efficiency is much less dependent on the motor's width.
Lastly we study the different characteristic length scales of the host medium in an attempt to find a match with the observed threshold values.

\section{Calculations}

\begin{figure}
\centering
  \includegraphics[height=4cm]{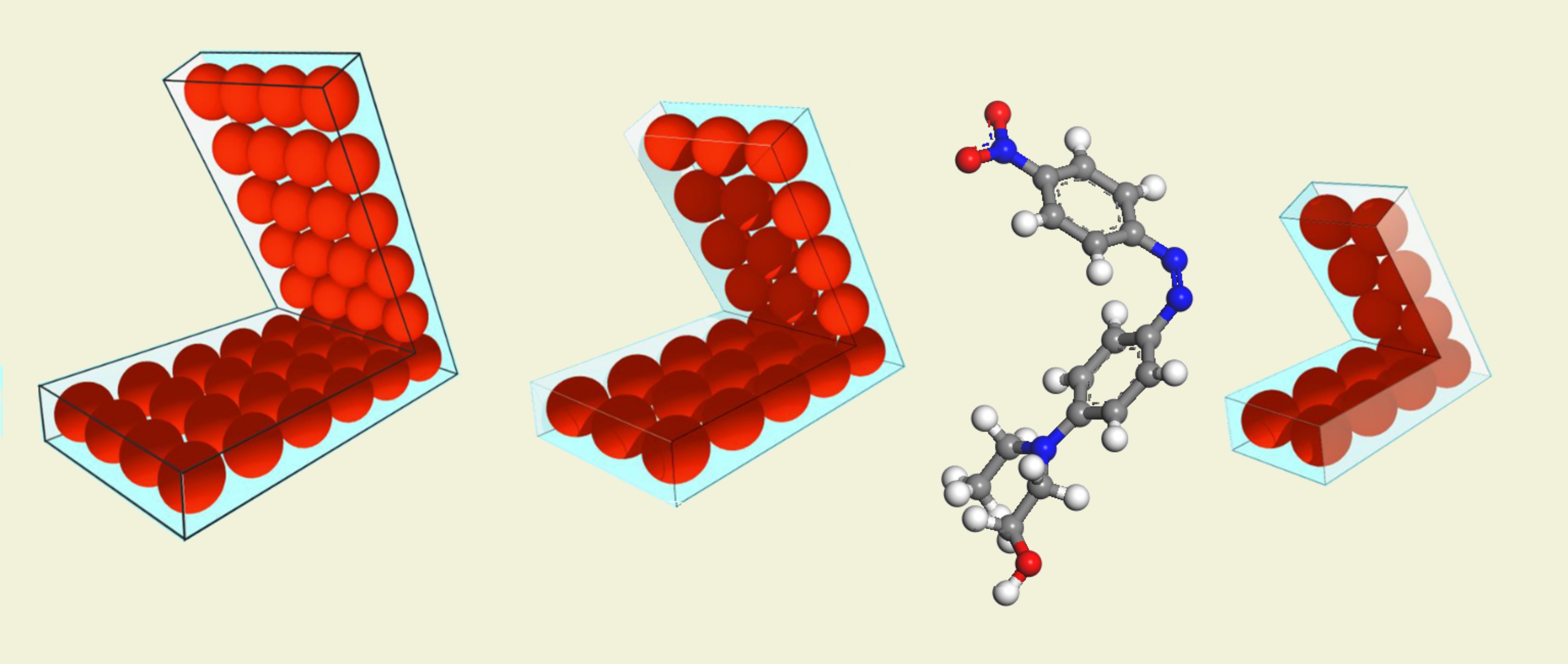} 
  \caption{ (color online) Picture of the closed (cis) forms of different model motor molecules constituted of parallel rows of atoms. 
  The angle between the two arms is $\theta=\pi/3$ in the model cis form. 
In the trans forms the molecules are flat. The real cis $DR1$ molecule is plotted for comparison. The rectangular lines are guides to the eye.\\}
\end{figure}

\begin{figure}
\centering
  \includegraphics[height=4cm]{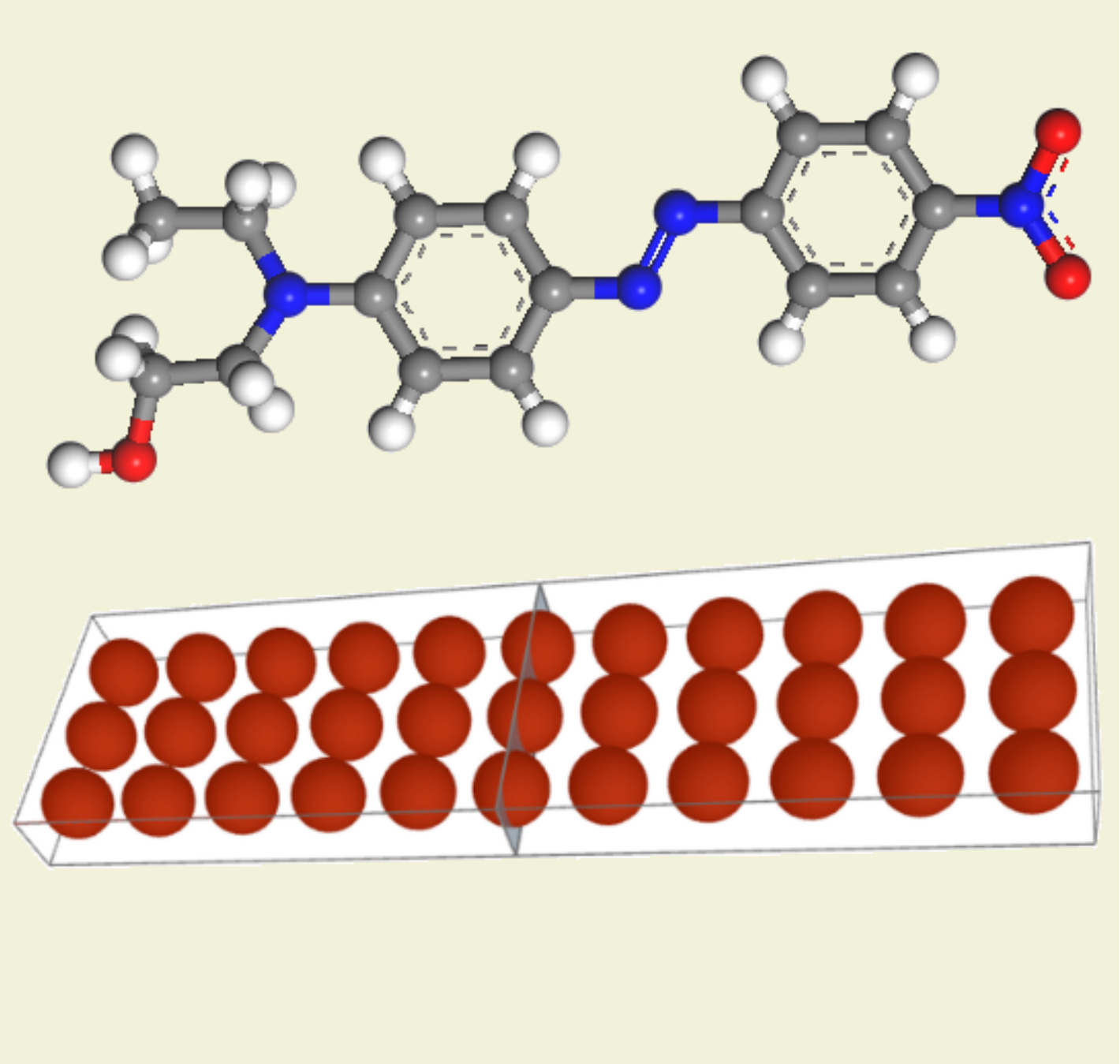}
  \caption{ (color online) Picture of the opened (trans) form of a model motor molecule constituted of parallel rows of atoms. 
In the trans forms the molecules are flat. The real trans $DR1$ molecule is plotted for comparison. Displayed lines are guides to the eye.\\}
\end{figure}

Our system contains $500$ diatomic host molecules and one molecular motor diluted  inside the host medium. 
We use different host molecules each constituted of two atoms ($i=1, 2$) that do interact with  Lennard-Jones potentials and that differ from the parameter $\alpha$.
$V_{ij}=4\epsilon_{ij}((\sigma_{ij}/r)^{12} -(\sigma_{ij}/r)^{6})$ with the parameters\cite{ariane}: $\epsilon_{11}= \epsilon_{12}=0.5 KJ/mol$, $\epsilon_{22}= 0.4 KJ/mol$, $\sigma_{ij}= \alpha \sigma_{ij}^{0}$ where $\sigma_{11}^{0}= \sigma_{12}^{0}=3.45$\AA, $\sigma_{22}^{0}=3.28$\AA$ $ and $\alpha$ is a constant that defines our different hosts.
We use the mass of Argon for each atom of the linear host molecule that we rigidly bond fixing the interatomic distance to $d=\alpha.d^{0}$ with $d^{0}=1.73 $\AA$ $.  
With these parameters the host (alone or mixed with the probe) does not crystallize even during long simulation runs\cite{ariane}.
We model the motor with a rectangular shape constituted of rows of Lennard-Jones atoms with parameters: 
$\epsilon_{33}= 0.996 KJ/mol$,  $\sigma_{33}=3.405$\AA. 
We use the following mixing rules\cite{md1}: $\epsilon_{ij}=(\epsilon_{ii} . \epsilon_{jj})^{0.5}$ and $\sigma_{ij}=(\sigma_{ii} . \sigma_{jj})^{0.5}$  for the interactions between the motor and the host atoms. As for the host, we use the mass of Argon for each atom of the motor.
The distance between atoms varies for different motors but we chose the distance to be smaller than  $\sigma_{33}$ to obtain a continuous molecule.
Unless otherwise specified, we chose a motor with a length  $L_{T}$=15.4 \AA\ and width W=3.9 \AA.
 The rectangular and flat motor folds periodically, with two symetrical arms as shown in Figures 1 and 2.\\
With these parameters, below $T=38 K$ the system falls out of equilibrium in our simulations, i.e. $T=38$ K is the smallest temperature for which we can equilibrate the system while the chromophore does not isomerize. 
As a result above that temperature the medium behaves as a viscous supercooled liquid in our simulations and below that temperature it behaves as a solid (as $t_{simulation}<\tau_{\alpha}$). We evaluate the glass transition temperature $T_{g}$ to be slightly smaller  $T_{g} \approx 28 K$, from the change of the slope of the potential energy as a function of the temperature.  
 However as they are modelled with Lennard-Jones atoms, the host and motor potentials are quite versatile.
Due to that property, a shift in the parameters $\epsilon$ will shift all the temperatures by the same amount, including the glass-transition temperature and the melting temperature of the material. We have used mainly two different hosts that we will call host1 ($\alpha=1$) and host2 ($\alpha= 1.32$).\\ 

We use the Gear algorithm with the quaternion method\cite{md1} to solve the equations of motions with a time step $\Delta t=10^{-15} s$. 
The temperature is controlled using a Berendsen thermostat\cite{berendsen}. We use periodic boundary conditions.
When the motor is not activated, the simulation box size doesn't affect the results as long as the number of host molecules is significantly larger than $N_{host}=100$.
When the motor is activated, an increase of the simulation box size corresponds to a decrease of the concentration of motor molecules, leading to a decrease of the average induced diffusion.  However the diffusion in the vicinity of the motor is not affected by an increase of the box size. In agreement with that result, we found that the effect of the motor on the host decreases exponentially with the distance, leading to no long range effects. 
We model the isomerization as a uniform closing and opening of the probe molecule shape\cite{prl,cage,ivt4}. The period $\tau_{p}$  of the isomerization $cis-trans$ and then $trans-cis$ is also fixed in each study ($\tau_{p}=1 ns$ (host1) or $600ps$ (host2)). During the isomerization, the shape of the motor molecule is modified slightly at each time step using a constant step quaternion variation. That quaternion step is calculated so that the molecule is in the final configuration (folded or unfolded) at the end of the isomerization.
 The isomerization time is set at $t_{iso}=1 ps$, a value chosen from the typical DR1 folding time. 
The molecule thus folds continuously during $t_{iso}$ then stays in the $cis$ shape during $\tau_{p}/2 - t_{iso}\approx  \tau_{p}/2$ then unfolds continuously during $t_{iso}$, then stays in the $trans$ shape during $\tau_{p}/2 - t_{iso}$ and the cycle begins again.
The foldings  take place at periodic intervals whatever the surrounding local viscosity.
This approximation has recently been validated experimentally\cite{diamond} as the pressure that is necessary to stop the azobenzene isomerization is very large ($P> 1$  $GPa$).

\begin{figure}
\centering
  \includegraphics[height=6cm]{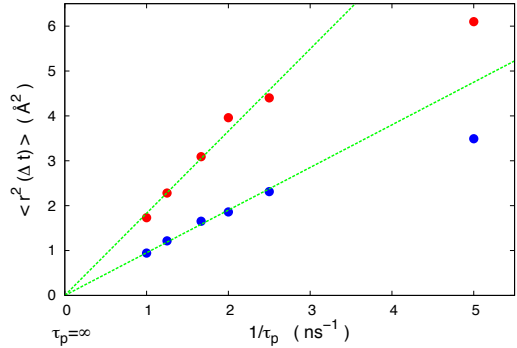}
  \caption{(color online) Host mobility $<r^{2}(\Delta t)>$ ($\Delta t =10$ ns) dependence on the period $\tau_{p}$ between two trans-cis isomerizations of the motor.
  The small thermal mobility has been subtracted from the results. From top to bottom: Red circles (host 1):  $\alpha=1$, $\sigma_{11}= 3.45 $ \AA, $\rho=2.42 g/cm^{3}$,  $T=10 K$ $ (T<T_{g})$;  Blue circles (host 2): $\alpha=1.32$, $\sigma_{11}= 4.56 $ \AA, $\rho=1.63 g/cm^{3}$, $T=80 K$ $(T>T_{g})$. The motor has a length  $L_{T}$=15.4 \AA\ and width W=3.9 \AA.
  According to the linear response theory, the response is proportional to the stimulus, provided that the stimulus is small enough.
  We see that we are in that regime for $1/\tau_{p} \leq 2.5 ns^{-1}$ i.e. $\tau_{p} \geq 0.4  ns$, while for larger frequencies a saturation appears due to the decrease in the viscosity of the medium around the motor.
  \\}
\end{figure}

In a previous work\cite{ivt4} we studied  in detail the effects of the isomerization period $\tau_{p}$ on our results.
We found that with this large period an isomerization doesn't influence the behavior of the system long enough to affect the next isomerization effect. That means that our period is large enough for each isomerization to be an independent process.
However experiments using azobenzene isomerizations often use very small light intensities, resulting to a much larger mean period of time between isomerizations than can be achieved with simulations. It is thus important to check in our particular model that we are in the same physical regime than in experiments. 
Within the linear response theory, the response is proportional to the stimulus provided that the stimulus is small enough and the period between two stimuli is large enough.
If we are inside the linear response regime with our time periods, the larger time periods (or smaller stimuli) used in most experiments will induce the same regime. 
We show in Figure 3 that the diffusion of host molecules is proportional to the inverse of the period in our model, up to a frequency $f= 1/\tau_{p}=2.5$ $ ns^{-1}$ i.e. for  periods $\tau_{p} \geq 400 ps$. As a result in our study ($\tau \geq 600 ps$ i.e. $1/\tau_{p} \leq1.66$ $ns^{-1}$ ) we are located inside the linear response regime, that is the regime of most experiments.\\
Finally, when the two arms of our model motor fold, the distance covered by each arm is $\Delta l=L\theta'=1.05 L$. ($\theta'=\pi/2-\theta/2=\theta=\pi/3$ and $L$ is the arm's length). 
Nearby host molecules can thus be pushed during the folding to a maximum distance $\Delta l=1.05L$.

\section{Results and discussion}

\subsection{Comparison between the motions induced by the simplified motors and by the azobenzene molecule}

We begin our study with a comparison of the effect of our simplified motor and of the real $DR1$ molecule on the surrounding host material's diffusion. Our purpose here is to validate the results that we will obtain with the simplified motors, but also to investigate the pertinent motions and shape of the motor to induce diffusion.
We show in Figure 4 that comparison for the mean square displacements (MSD) of the hosts molecules surrounding the motor,  for the MSDs of the motor molecules, and for the Van Hove correlation functions of the host.   

\begin{figure}
\centering
  \includegraphics[height=6cm]{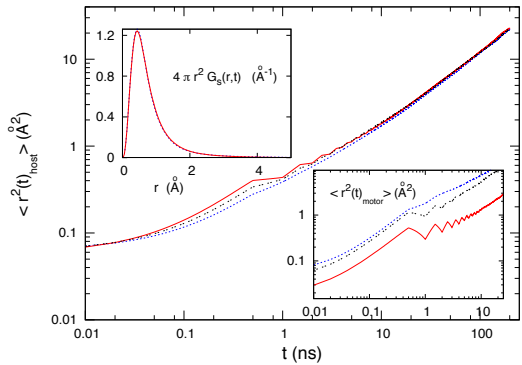}
  \caption{(color online) Comparison of the mean square displacements $<r^{2}(t)_{host}>$ of the host molecules (host 1, $\sigma=3.45$\AA, $l=5.1$\AA, T=10 K) when the motor is the $DR1$ molecule (red continuous curve), a model motor with a length  $L_{T}$=2*6.2=12.4 \AA\ and width W=5.4 \AA (black dashed curve), or a model motor with a length  $L_{T}$=2*5.7=11.4 \AA\ and width W=3.9 \AA\  (blue dotted curve).The periodic oscillations in the mean square displacements correspond to the isomerization period. These oscillations show that periodically the isomerization of the motor pushes the host molecules away, increasing the host molecules displacements.
 Inset  (at the bottom): Mean square displacements $<r^{2}(t)_{motor}>$ of the motor molecule in the same conditions than above.
  As there is only one motor in the simulation box, these results are much less accurate than for the host.
The oscillations are here again due to the periodic isomerizations of the motor molecule. 
These oscillations show that the isomerization modifies the motor's motion inside the host medium.
Upper inset: Comparison of the Van Hove correlation functions $G_{s}(r,t)$ ($t=400 ps$) of the host molecules. 
}
\end{figure}

We find  similar isomerization-induced host diffusive motions using the real DR1 motor molecule or our simplified one. 
For the three different MSD curves, the host molecules begin escaping the cages created by their neighbors around $0.02$ ns (ending of the plateau of the MSD) and reach the same diffusive behavior for $t>10$ ns.
This is quite surprising to find similar isomerization-induced motions for the three curves as the shapes and closing processes are quite different.  
This result suggests that the detailed shape of the motor is not of major importance to induce diffusion  inside the host medium, because different shapes lead to the same diffusion.
The model motor isomerizes in a simple folding, 
however that simple folding leads to similar induced diffusive motions than the complex folding of the DR1 molecule. 
That result suggests that the host motions are mainly induced by the perpendicular closing of the chromophore and that the complex closing of the $DR1$ molecule is not necessary to induce diffusion inside the host medium.
The upper inset compares the Van Hove distribution functions that represent the probability to find a molecule at time t a distance $r$ apart from its initial position.
We find similar Van Hove distribution functions with the simplified motor and with the $DR1$ molecule with a small timescale difference in agreement with the MSD results.

If the host diffusion is the same for the three motors of Figure 4,  the motor diffusion is not the same. The diffusion of the motor thus depends on its shape. Note that as we use logarithmic scales in the Figure,  the diffusion coefficient $D$ is not related to the slope of the mean square displacement,  as it will be for a linear scale, but to the height of the curves. 
Figure 4 shows that the diffusion of the motor is significantly larger for the two model motors displayed than for the true $DR1$ molecule (red continuous curve).
Interestingly, the larger motor (black curve) is not the most efficient in the Figure.
These results suggest that it is possible to design motors that move faster inside the host material\cite{compare} and that interesting possibility will be the subject of further work.\\

To conclude this first part of our study, we find that the detailed closing of the $DR1$ molecule as well as it's precise shape are not necessary to induce diffusive motions  inside the host medium. We also find that the isomerization induced diffusion arises mainly from the perpendicular closing of the motor molecule and not from the motor arms partial rotation that takes place during that closing. These results are in agreement with the versatility of the chromophore's ado-dyes that experimentally induce surface relief gratings on various hosts.\\

\subsection{Molecular motions dependence on the motor's shape}

\subsubsection{Motor's length effect}

The host diffusion coefficient variation with the arm length displays in Figure 5  a threshold followed by a linear increase of the diffusion. 
We interpret the linear increase in the framework of the linear response theory, as a consequence of a small increasing stimulus. The presence of thresholds agrees well with at least two theories of the isomerization-induced massive transport. In the cage breaking theory the host molecules have to be pushed at least to a nearby cage to induce motion, leading to a minimum arm length. In the pressure gradient theory the induced pressure gradient has to overcome the constitutive stress of the material, leading also to a minimum arm length to initiate motion.
Figure 6 shows that the inverse of the relaxation time\cite{relaxation} $1/\tau_{\alpha}$ presents a threshold as well, followed by a linear increase with respect to the motor's arm length $L$.
Since the relaxation time is related to the viscosity of the material, that result shows that the viscosity decreases (intermittently) around the motor only above a threshold value. Below that value of $L$, the viscosity of the host material is thus unaffected by the motor's motions in our simulations. 

The motor's arm length threshold is $L_{threshold}\approx 5$\AA\ in these results. Note that the host molecule corresponding to these data is $w=3.4$\AA\ wide and $l=5.1$\AA\ long.
The threshold is thus roughly equal to the host molecule's size.

\begin{figure}
\centering
\includegraphics[height=6cm]{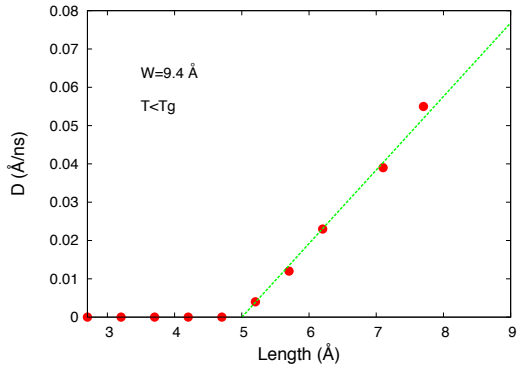}
\caption{(color online) Host diffusion coefficient versus the length of the arm of the molecular motor ($L=L_{T}/2$) for a constant motor's width W=9.4 \AA.
(host 1, $\sigma=3.45$\AA, $l=5.1$\AA, T=10 K)\\}
\end{figure}

\begin{figure}
\centering
\includegraphics[height=6cm]{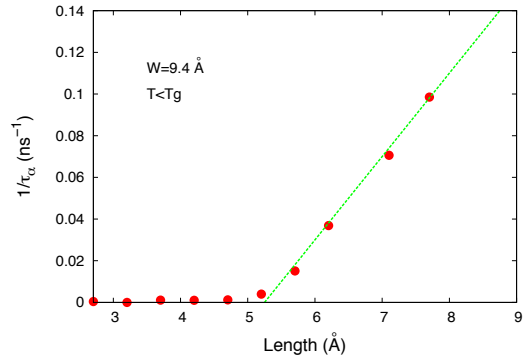}
\caption{(color online) Average $\alpha$ relaxation time ($\tau_{\alpha}$) of host molecules inside the simulation box versus the length of the arm of the molecular motor ($L=L_{T}/2$) for a constant motor's width W=9.4 \AA.
(host 1, $\sigma=3.45$\AA, $l=5.1$\AA, T=10 K)\\}
\end{figure}

\begin{figure}
\centering
\includegraphics[height=6cm]{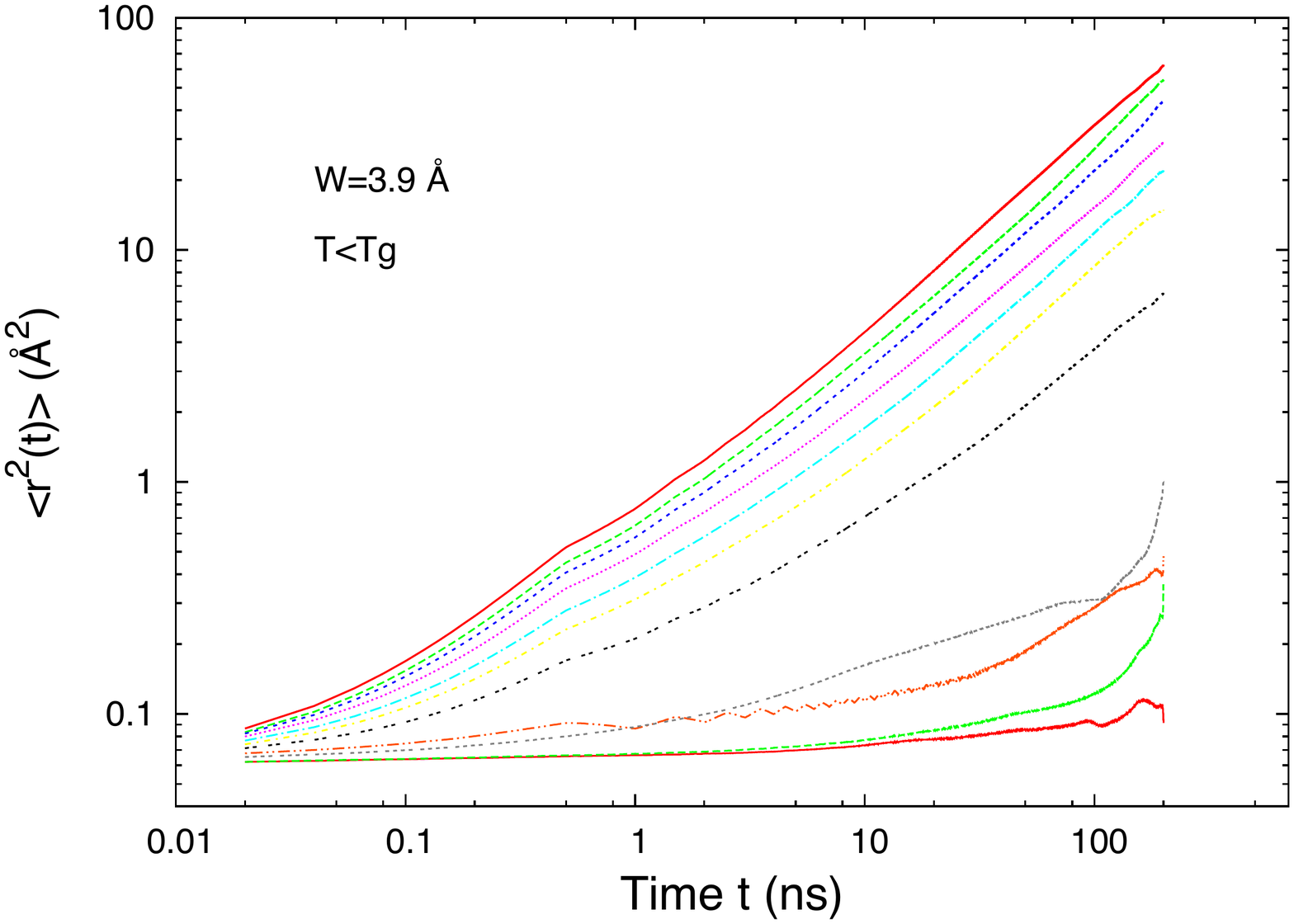}
\caption{(color online) Mean square displacement of host 1 molecules centres of masses below Tg when the isomerization is set on.
The different curves represent different chromophores lengths. From bottom to top (right hand side) $L_{T}=$5.4, 6.4, 7.4, 8.4, 9.4, 10.4, 11.4, 12.4, 14.4, 15.4, and 17.4 \AA.
The diffusive motions begin for $L_{T}=9.4$ \AA, i.e. $L=L_{T}/2=4.7$ \AA. (host 1, $\sigma=3.45$\AA, $l=5.1$\AA, T=10 K)\\}
\end{figure}

The mean square displacements (MSD) behavior, displayed in Figure 7 illustrates the threshold appearance. 
Below the threshold, the motions oscillate due to the periodic isomerizations but without reaching diffusion that would lead to a MSD that increases linearly with time according to the Stokes-Einstein law $<r^{2}(t)>=6Dt$ for Brownian motion.
Figure 7 also shows that above the threshold (for $L>5$\AA) the molecular motions are diffusive on long times.
We find the same behavior at higher temperatures (T=40 K), however in this case the diffusion never cancels totally due to the presence of thermal diffusion. The thermal diffusion has thus to be subtracted from our higher temperatures values to observe the threshold.\\

\subsubsection{Motor's width effect}

\begin{figure}
\centering
  \includegraphics[height=6cm]{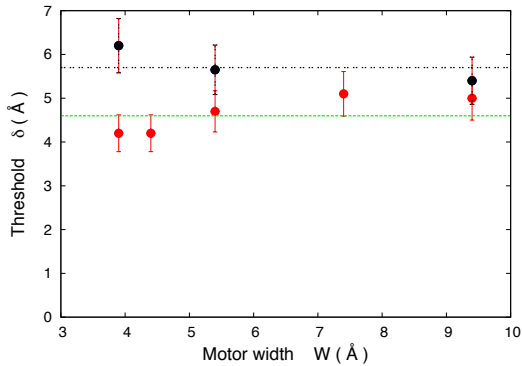}
  \caption{(color online) Threshold on the motor's length versus the width of the motor. The different symbols correspond to two different hosts: Red circles: host 1, T=10K; Black circles: host 2, T=40K.
  The results show that the width dependence of the threshold is slight.\\}
\end{figure}

In the cage breaking picture we do not expect any threshold for the width as increasing the arm width does not change the distance we are able to push the surrounding molecules but only the number of molecules that can be pushed away.
In agreement with that picture we didn't find any width threshold for the diffusion.
If we do not expect any width threshold, we expect that multiplying the width by a mere factor $2$ is roughly equivalent as having two chromophores instead of only one. 
In that picture, in the linear response regime the diffusion must be proportional to the width of the chromophore's arm and we expect a rough linear increase of the diffusion with the width.  But our results are in contradiction with that picture, as Figure 8   shows that the diffusion is almost constant whatever the chromophore's width.
That result suggests that the chromophore's motion disturbs the motion of molecules in its vicinity resulting in a reduction of the wider motor ability to push more molecules.

\subsection{Characteristic length scales that match the threshold values}

In a general picture, the presence of a motor length threshold for the induced diffusion indicates the presence of a characteristic length scale in the medium that needs to be surpassed for efficient induced motions to occur.
To determine the length scale involved, we  investigate the different characteristic length scales of the host material and compare them with the threshold value. For more clarity we make use here of host1 unless otherwise mentioned. For host 1 we found a motor arm's threshold of $4.8$\AA\ ${\pm} 0.5$\AA, leading to an arm displacement threshold $l\theta'= 5$\AA\  ${\pm} 0.5$\AA.

In supercooled liquids and amorphous materials, molecules are transiently trapped inside the cage constituted by their neighbors and the length scale of  importance in free-volume theories is the distance $L_{cage}^{host}$ of free motions inside the cage. From the plateau of the mean square displacement at low temperature $<r^{2}_{plateau}>=0.06 $\AA$^{2}$ for the host molecules, we obtain $L_{cage}^{host}=\sqrt{<r^{2}_{plateau}>}=0.25$\AA. 
Similarly, for the motor  $<r^{2}_{plateau}>=0.94 $\AA$^{2}$  leads to $L_{cage}^{motor}=0.97$\AA.
These lengths (0.25\AA\ and 0.94\AA) are much smaller than our threshold values that range in between 4 and 5 \AA.
As a result the explanation of the induced diffusion is not to be searched in the free volume modifications that occur during the isomerizations.

A second characteristic length scale is given by the size of the host molecule. The host 1 molecule is linear with a length of $5.1$\AA\ and a maximal width of $3.45$\AA. The length of the molecule agrees thus remarkably well for host 1 with the arm displacement threshold. This result is in qualitative agreement with the cage-breaking picture, as the size of the host molecule in that picture governs the possibility to induce diffusion. To be more precise however we calculate the radial distribution function $g(r)$ between host molecules (Figure 9).
$g(r)$ represents the probability density to find a molecule a distance $r$ apart from another molecule. This function gives information on the microscopic structure of the material. The $g(r)$ first peak location corresponds to the distance between a molecule and its first neighbors, while the second peak corresponds to the second neighbors location.
We find in Figure 9 a first peak located at $r_{0}=3.7$\AA\ followed by a second peak at $r_{1}=4.4$\AA.
Then the first minimum is located at $r_{min}=5.7$\AA. This last value ($r_{min}$) is the length scale of importance for cage breaking mechanisms as it corresponds to the location of the barrier of the mean force potential\cite{md1} $V_{mf}=-kT . log( g(r) )$.
The distance to the potential barrier is the distance that must be overpassed by a host molecule to escape the cage of its neighbors, and thus the distance the motor must push  host molecules away to induce diffusion in the cage breaking model.

\begin{figure}
\centering
\includegraphics[height=6cm]{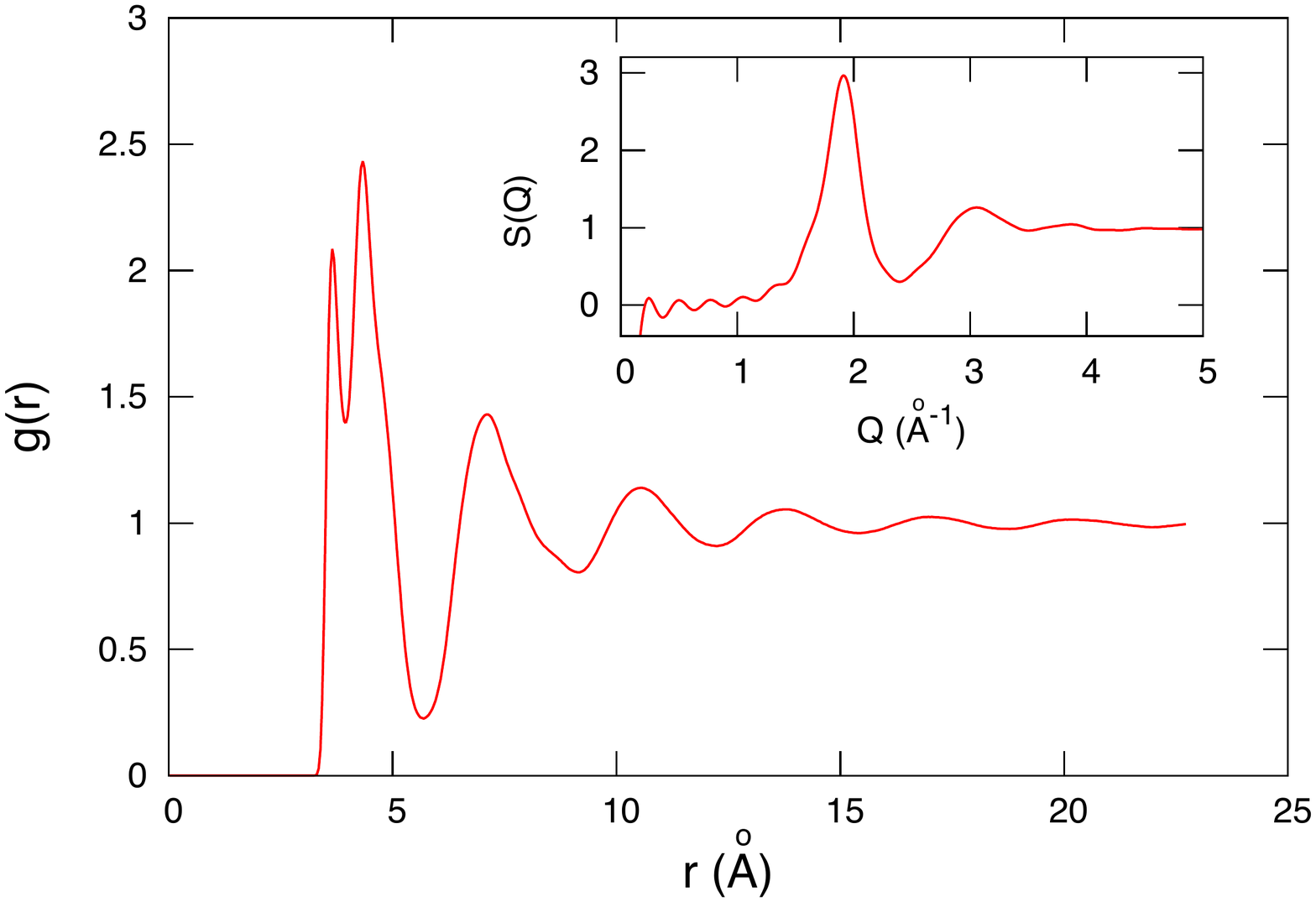}
\caption{(color online) Radial distribution function $g(r)$ between centers of mass of host 1 molecules. T=10K. Inset: Structure factor S(Q). The first peak of the structure factor is located at $Q_{0}=1.92$\AA$^{-1}$ leading to a global correlation length $L_{global}=2\pi/Q_{0}=3.3$\AA.\\}
\end{figure}

The size of the molecule $5.1$\AA\ and the position of the first minimum of the radial distribution function $r_{min}=5.7$\AA\ correspond approximately to the observed arm displacement threshold value $l\theta'= 5$\AA.
These two lengths (the molecule's size and  $r_{min}$) are directly related, but the size of the molecule is of particular interest for experimental purposes as its value is usually known without having to make any computation.

\begin{figure}
\centering
  \includegraphics[height=6cm]{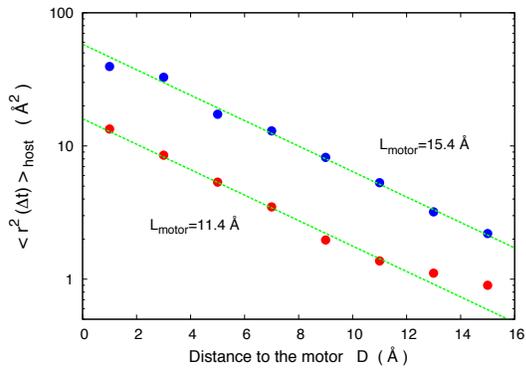}
  \caption{(color online) Local host mobility $\mu=<r^{2}(\Delta t)>$ ($\Delta t=10 ns$) as a function of the distance $D$ from the motor, for two different motor molecules.  
 The curves display an exponential decrease $\mu=\mu_{0} e^{-D/R_{0}}$ with $R_{0}=4.55$ \AA\ that is reminiscent of the effect of a wall on the mobility of confined supercooled liquids\cite{confit} where it is attributed to cooperative mechanisms. The host size is $\sigma=4.56$\AA. Host 2, T=40K, $\alpha=1.32$.}
\end{figure}

The two models that predict thresholds, (the gradient pressure model\cite{a22,a23}and the cage breaking model\cite{cage}) are difficult to distinguish due to their similarities.
To investigate the gradient pressure model,
in Figure 10 we display the variation of the induced mobility in the host (host 2) with the distance of the motor. For this calculation we divide the sphere surrounding the motor in slices of 2 \AA\ width and calculate the average host mobility inside each slice. We find that the mobility of the host decreases exponentially with the distance to the motor $\mu=\mu_{0} e^{-D/R_{0}}$ with $R_{0}=4.55$ \AA. The host size is here $\sigma=4.56$\AA\,  i.e. $ R_{0}\approx \sigma$. 
In the gradient pressure model we would have expected an $1/4\pi r^{2}$ decrease of the diffusion, due to the solid angle decrease of the stimulus (the gradient pressure), instead of that exponential decrease. In that model we would also have expected a distance threshold for diffusion (i.e. that the diffusion stops at distances $r>r_{threshold}$), because in that model  the gradient pressure (that decreases with the distance to the force induced by the motor's isomerization on nearby molecules) must overpass the constitutive constraint of the material to induce motions.  

\subsection{Threshold dependence on the host molecule size}

In the previous subsection we found that the threshold $\delta$ was approximately equal to the size of the host molecule.
To test that relation further we will vary the size of the host and study the induced modification of the threshold. 
In Figure 11 we use the real DR1 as motor molecule and vary the size of the host by multiplying the parameters $\sigma_{ij}$ of the Lennard Jones potential, and the distance between the two atoms of the molecule by the same factor $\alpha$.
We find that the host diffusive motions stop when the width of the host  $\sigma>5.4$ \AA\  i.e. the length of the host  $l>8$\AA.
Interestingly enough that value corresponds approximately to the lightest arm's (the arm that moves the most) size of the DR1 molecule: $l_{arm}=7.8$\AA.
These results confirm that the threshold is controlled by the size of the host molecule, as using a constant motor's arm the motion stops when the host size is too large.

In Figure 12, using our model motors, we vary the host molecule size to study its effect on the threshold value.
Each point on that plot correspond to a different host for which we varied the motor's length to find the threshold.
The radial distribution function minimum value $r_{min}$ and the length of the molecule both evolve here proportionally to the size $\sigma$.  
Figure 12 shows that the threshold value is proportional to the size of the molecule in the conditions of our study.
This result confirms the existence of a direct relation between the threshold value and the host molecule size.

\begin{figure}
\centering
  \includegraphics[height=6cm]{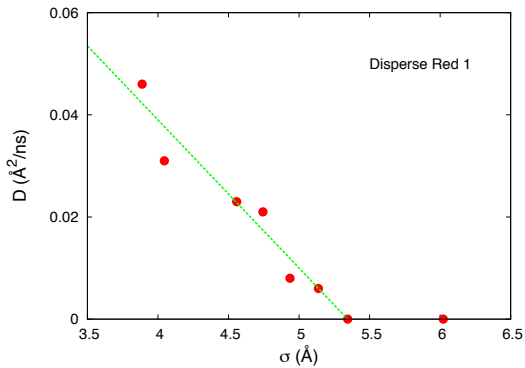}
  \caption{(color online) Host diffusion coefficient versus the size of the host molecule, for a constant motor (the real DR1). $\sigma=\alpha.\sigma^{0}$ is here the size parameter $\sigma_{11}$ in the Lennard Jones potential between atoms of type $1$. We chose $\sigma_{22}$ and the distance between the two atoms to increase proportionally ($d=\alpha d^{0}$ and $\sigma_{22}=\alpha. \sigma_{22}^{0}$), so that $\sigma$ varies proportionally to the length $l=d+(\sigma_{11}+\sigma_{22})/2$ and width $w=(\sigma_{11}+\sigma_{22})/2$ of the host molecule. Thus $l=1.48.\sigma$ and $w=0.98.\sigma$. The size of the simulation box is also rescaled in the same way.\\}
\end{figure}

\begin{figure}
\centering
  \includegraphics[height=6cm]{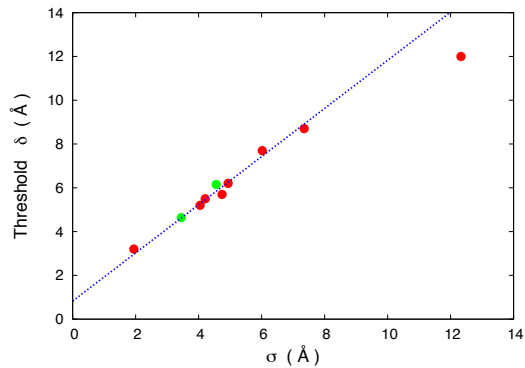}
  \caption{(color online) Threshold values $\delta$ on the motor's length versus the size ($\sigma$) of hosts molecules. The host molecules are defined by the value of $\sigma$. $\sigma=\sigma_{11}$, $l=1.48 \sigma$ and $w=0.98 \sigma$.   The green (light) circles correspond to thresholds obtained from simulations at T=10K and the red (dark) circles  T=40K. The line in the Figure corresponds to a threshold $\delta=1.1 \sigma +0.75$\AA. The results strongly suggest a  threshold $\delta$ determined by the size of the host molecules. The last point to the right is affected by a larger uncertainty due to the large size of the host molecules in comparison to the motor's atoms.\\}
\end{figure}

\section{Conclusion}
The aim of this work was to find the significant parameters that control the motion of the molecules around the folding motor. 
We found motors size thresholds for the motions of the molecules. That behavior appears for the real DR1 motor molecule as well as with our simplified folding motor. In both cases the threshold is directly related to the size of the host molecule. The size of the motor relative to the size of the host is thus a relevant parameter for the massive mass transport mechanism, in good agreement with the cage-breaking model\cite{cage}.

The thresholds do not depend significantly on the width of the motor, also in agreement with the model.
The transport increases linearly with the length of the motor, a result that shows that we are in the linear theory regime, and is related to an increase of the number of molecules pushed by the motor. That increase of the transport with the motor's size is in agreement with recent experiments\cite{exper}.
We found that the induced motions decrease exponentially with the distance to the motor. The same behavior has been observed in confined supercooled liquids\cite{confit} where it is attributed to cooperative motions.
Results show that the DR1 molecule and simplified motors of similar sizes, lead to similar transport, a result that shows that the system is versatile and explains in part the universality of the observed effects.
However the shape of the motor influences its own motion relative to the host. This last result could permit to create new motors that have the ability to move efficiently inside the medium and work is in progress in that direction.

\vskip 0.5cm
{\bf Aknowledgments:}

We would like to thank Samuel Migirditch from Appalachian State University for his help and discussions at the end of this work and professor Jean-Luc Fillaut from Rennes University for interesting discussions at the beginning of this work.


\vskip 0.5cm


\begin{thebibliography}{99}


\bibitem{motor1} R.D. Astumian, 
 \newblock \emph{Science } {1997}, {\bf 276}, 917-922.

\bibitem{motor2} M.F. Hawthorne, J.I. Zink, J.M. Skelton, M.J. Bayer, C. Liu, E. Livshits, R. Baer, D. Neuhauser, 
 \newblock \emph{Science } {2004}, {\bf 303}, 1849-1851.

\bibitem{motor3} P. Palffy-Muhoray, T. Kosa, E. Weinan, 
 \newblock \emph{Appl. Phys. A} {2002}, {\bf 75}, 293-300.

\bibitem{motor4} J. Berna, D.A. Leigh, M. Lubomska, S.M. Mendoza, E.M. Perez, P. Rudolf, G. Teobaldi, F. Zerbetto, 
 \newblock \emph{Nature Mater.} { 2005}, {\bf 4}, 704-710.

\bibitem{motor5} T.R. Kline, W.F. Paxton, T.E. Mallouk, S. Ayusman,
 \newblock \emph{Angew. Chem. Int. Ed.} { 2005}, {\bf 44}, 744-746.

\bibitem{motor6} W.R. Browne, B.L. Feringa, 
 \newblock \emph{Nature Nanotech.} { 2006}, {\bf 1}, 25-35.

\bibitem{motor7} K. Dholakia, P. Reece, 
 \newblock \emph{Nanotoday} { 2006}, {\bf 1}, 20-27.


\bibitem{motor8} T. Fehrentz, M. Schonberger, D. Trauner, 
 \newblock \emph{Angew. Chem. Int. Ed.} { 2011}, {\bf 50}, 12156-12182.


\bibitem{motor10} M.M. Russew, S. Hecht,
 \newblock \emph{Adv. Mater.} { 2010}, {\bf 22}, 3348-3360.

\bibitem{motor11} N. Katsonis, M. Lubomska, M.M. Pollard, B.L. Fearing, P. Rudolf,
 \newblock \emph{Progress in Surface Science} { 2007}, {\bf 82}, 407-434.


\bibitem{motor12} A.P. Davis,
 \newblock \emph{Nature} {1999}, {\bf 401}, 120-121.

\bibitem{motor13} J.P. Sauvage,
 \newblock \emph{Molecular machines and motors} { 2001}, {Springer, Berlin}.


\bibitem{motor14} E.R. Kay, D.A. Leigh,
 \newblock \emph{Nature} {2006}, {\bf 440}, 286-287.


\bibitem{motor15} V. Balzani, et al.,
 \newblock \emph{Proc. Nat. Acad. Sci.} {2006}, {\bf 103}, 1178-1183.


\bibitem{motor16} T. Muraoka, K. Kinbara, Y. Kobayashi, T. Aida,
 \newblock \emph{JACS} {2003}, {\bf 125}, 5612-5613.


\bibitem{motor17} T.J. Huang, et al.,
 \newblock \emph{Nano Lett.} {2004}, {\bf 4}, 2065-2071.

\bibitem{review} A. Natansohn,  P. Rochon, 
 \newblock \emph{Chem. Rev. } {  2002}, {\bf102}, 4139-4175.
 

\bibitem{review2} J.A. Delaire, K. Nakatani, 
 \newblock \emph{ Chem. Rev.} {  2000}, {\bf 100}, 1817.
 

\bibitem{review4} K.G. Yager, C.J. Barrett,  
\newblock \emph{Curr. Opin. Solid State Mater. Sci.} {  2001},
{\bf 5}, 487.



\bibitem{a17}  T.G. Pedersen, P.M.  Johansen, 
\newblock  {\em Phys. Rev. Lett.} {  1997}, {\bf  79}, 2470-2473.

\bibitem{a18}  T.G. Pedersen, P.M.  Johansen, N.C.R.  Holme, P.S.  Ramanujam, 
\newblock  {\em Phys. Rev. Lett.} {  1998}, {\bf  80}, 89-92.

\bibitem{a21}  J. Kumar, L.  Li, X.L.  Jiang, D.Y.  Kim, T.S.  Lee, S. Tripathy, 
\newblock  {\em Appl. Phys. Lett.} {  1998}, {\bf  72}, 2096-2098.



\bibitem{a22}  C.J. Barrett, P.L.  Rochon, A.L.  Natansohn, 
\newblock  {\em J. Chem. Phys.} {  1998}, {\bf  109}, 1505-1516.

\bibitem{a23}   C.J. Barrett, A.L. Natansohn, P.L. Rochon, 
\newblock  {\em J. Phys. Chem.} {  1996}, {\bf  100}, 8836-8842.

\bibitem{a24}  P. Lefin, C.  Fiorini,  J.M. Nunzi, 
\newblock  {\em Pure Appl. Opt.} {  1998}, {\bf  7}, 71-82.



\bibitem{diamond}  T.A. Singleton, K.S. Ramsay,  M.M. Barsan, I.S. Butler, C.J. Barrett, 
\emph{J. Phys. Chem. B} {  2012}, {\bf116}, 9860-9865.



\bibitem{dif1} P. Karageorgiev, D. Neher, B. Schulz, B. Stiller, U. Pietsch, M. Giersig, L. Brehmer, 
\emph{Nature Mater.} {  2005}, {\bf4}, 699-703.

\bibitem{dif2} G.J. Fang, J.E. Maclennan, Y. Yi, M.A. Glaser, M. Farrow, E. Korblova, D.M. Walba, T.E. Furtak, N.A. Clark, 
\emph{Nature Comm.} {  2013}, {\bf4}, 1521.

\bibitem{dif3} N. Hurduc, B.C. Donose, A. Macovei, C. Paius, C. Ibanescu, D. Scutaru, M. Hamel, N. Branza-Nichita, L. Rocha,
\emph{Soft Mat.} {  2014}, {\bf 10}, 4640-4647.

\bibitem{dif4}  J. Vapaavuori, A. Laventure, C.G. Bazuin, O. Lebel, C. Pellerin, 
\emph{J. Amer. Chem. Soc.} {  2015}, {\bf 137}, 13510.


\bibitem{regis}  R. Barille, S. Ahmadi-Kandjani, S. Kurcharski, J.M.  Nunzi, 
\newblock  {\em Phys. Rev. Lett. }, { 2006},   {\bf 97}, 048701.

\bibitem{regis2} S. Ahmadi-Kandjani, R. Barille, S. Dabos-Seignon, J.M. Nunzi, E. Ortyl, S. Kucharski,   
\emph{Optics Letters}, { 2005}, {\bf 30}, 1986-1988.

\bibitem{regis3} R. Barille, S. Dabos-Seignon, J.M. Nunzi, S. Ahmadi-Kandjani, E. Ortyl, S. Kucharski, 
\emph{Opt. Express}, 2005, {\bf13(26)},10697-10702.



\bibitem{prl}  V. Teboul, M. Saiddine, J.M. Nunzi, 
\newblock  {{\em Phys. Rev. Lett. }} {  2009}, {\bf  103}, 265701.

\bibitem{cage}  V. Teboul, M. Saiddine, J.M. Nunzi, J.B. Accary, 
 \newblock \emph{J. Chem. Phys. } {  2011}, {\bf134}, 114517.
 

\bibitem{md1} M.P. Allen,   D.J. Tildesley, 
\newblock  {\em  Computer Simulation of Liquids}, Oxford University Press, New York 1990

\bibitem{md2} M. Griebel, S. Knapek, G. Zumbusch, 
\newblock  {\em  Numerical Simulation in Molecular Dynamics}, Springer-Verlag, Berlin 2007

\bibitem{md2b} D. Frenkel, B. Smit, 
\newblock  {\em  Understanding Molecular Simulation}, Academic Press, San Diego 1996




\bibitem{md3}  
C.F.E. Schroer, A. Heuer, 
\newblock  {\em Phys. Rev. Lett.} {  2013}, {\bf  110}, 067801.

\bibitem{md4}  
J.B. Accary, V. Teboul, 
\newblock  {\em J. Chem. Phys.} {  2012}, {\bf  136}, 0194502.


\bibitem{md6}  
A. Furukawa, K. Kim, S. Saito, H. Tanaka, 
\newblock  {\em Phys. Rev. Lett.} {  2009}, {\bf  102}, 016001.

\bibitem{md7}  
T. Iwashita, T. Egami, 
\newblock  {\em Phys. Rev. Lett.} {  2012}, {\bf  108}, 196001.

\bibitem{md8}  
T. Gleim, W. Kob, K. Binder, 
\newblock  {\em Phys. Rev. Lett.} {  1998}, {\bf  81}, 4404.

\bibitem{md9}  
E.J. Saltzman, K.S. Schweitzer, 
\newblock  {\em J. Chem. Phys.} {  2006}, {\bf  125}, 044509.


\bibitem{regis4}  H. Leblond, R. Barille, S.  Ahmadi-Kandjani, J.M. Nunzi, E. Ortyl, S. Kucharski, 
\emph{J. Phys. B: At. Mol. Opt. Phys.}, 2009, {\bf 42},205401.











\bibitem{a26}  F. Fabbri,  Y.  Lassailly, S.  Monaco, K. Lahlil, J.P. Boilot, J. Peretti, 
\newblock  {\em Phys. Rev. B} { 2012}, {\bf  86}, 115440.

\bibitem{a19}  V. Toshchevikov, M. Saphiannikova, G. Heinrich, 
\newblock  {\em J. Phys. Chem. B} { 2009}, {\bf  113}, 5032-5045.




\bibitem{ariane} A.P. Kerasidou, Y. Mauboussin, V. Teboul, 
 \newblock \emph{Chem. Phys.}, 2015, {\bf 450}, 91.




\bibitem{berendsen}  H.J.C. Berendsen,  J.P.M. Postma, W. Van Gunsteren,  A. DiNola, J.R.  Haak, 
 \newblock  {\em  J. Chem. Phys.} {\bf 1984}, {\bf 81}, 3684-3690.
 
 \bibitem{ivt4}   J.B. Accary, V. Teboul, 
\emph{J. Chem. Phys.} {\bf 2013}, {\bf139}, 034501.


\bibitem{compare}  M. Saiddine, V. Teboul,  J.M. Nunzi, 
 \newblock \emph{J. Chem. Phys. } {  2010}, {\bf133}, 044902.

\bibitem{relaxation} We obtain the $\alpha$ relaxation time $\tau_{\alpha}$ from the relation: $F_{s}(Q_{0},{\tau}_{\alpha})=e^{-1}$.
Where $Q_{0}=1.92$\AA$^{-1}$ is the wave vector corresponding to the maximum of the structure factor. $F_{s}(Q,t)$ is the self part of the intermediate scattering function 
$\displaystyle{F_{s}({\bf Q},t)={1\over N} Re( \sum_{i=1}^{n}e^{-i{\bf Q}.({\bf r}_{i}(t)-{\bf r}_{i}(0))}}) $. This function describes the autocorrelation of the density fluctuations at the wave vector ${\bf Q}$.


 \bibitem{angle}  V. Teboul,
\newblock  {\em J. Phys. Chem. B} {\bf 2015}, {\bf  119}, 3854.

 \bibitem{exper}  A. Goulet-Hanssens, T. C. Corkery, A. Priimagi, C. J. Barrett,
\newblock  {\em J. Mater. Chem. C} {\bf 2014}, {\bf  2}, 7505.


 \bibitem{confit}  V. Teboul, C. Alba-Simionesco
\newblock  {\em J. Phys.: Condens.Matter} {\bf 2002}, {\bf  14}, 5699-5709.










\end{thebibliography}
\end{document}